\documentclass[sigconf, nonacm]{acmart}

\usepackage{colortbl}
\usepackage{color}
\usepackage{xcolor}
\usepackage{caption}
\usepackage{subcaption}
\usepackage{framed}
\usepackage{makecell}
\usepackage{pifont}
\usepackage{amsmath}

\definecolor{keyaccent}{HTML}{0F3D7A}

\newcounter{insightcounter}

\newcommand{\insight}[2]{
  \vspace{3pt}
  \begin{framed}
  \refstepcounter{insightcounter}
  \noindent\mbox{{\bfseries Insight \theinsightcounter }\ | \textbf{#1}:}\hspace{0.5em}#2\par%
  \end{framed}
  \vspace{3pt}
}

\newcounter{recommendationcounter}

\newcommand{\recommendation}[2]{
  \vspace{3pt}
  \begin{framed}
  \refstepcounter{recommendationcounter}
  \noindent\mbox{{\bfseries Recommendation \therecommendationcounter }\ | \textbf{#1}:}\hspace{0.5em}#2\par%
  \end{framed}
  \vspace{3pt}
}

\newcommand{\parheader}[1]{%
  \par\addvspace{2mm plus 1mm minus 1mm}%
  \noindent\textbf{#1}%
}

\newcommand\vldbavailabilityurl{https://github.com/SvenHepkema/Valk/tree/arxiv-artifact}

\newcommand{\titlebreak}{}

\begin{document}
\title{Over the Memory Wall, Into the Instruction Wall: The New Bottleneck in GPU Data Processing}

\author{Sven Hepkema}
\affiliation{%
    \institution{Systems Group, ETH Zurich}
  \state{Switzerland}
}
\email{sven.hepkema@inf.ethz.ch}

\author{Bowen Wu}
\affiliation{%
  \institution{Systems Group, ETH Zurich}
  \state{Switzerland}
}
\email{bowen.wu@inf.ethz.ch}

\author{Christos Kozyrakis}
\affiliation{%
	\institution{NVIDIA \& Stanford University}
  \state{United States}
}
\email{ckozyrakis@nvidia.com}

\author{Yannis Chronis}
\affiliation{%
    \institution{Systems Group, ETH Zurich}
  \state{Switzerland}
}
\email{chronis@inf.ethz.ch}

\author{Gustavo Alonso}
\affiliation{%
    \institution{Systems Group, ETH Zurich}
  \state{Switzerland}
}
\email{alonso@inf.ethz.ch}

\begin{abstract}
	Datacenter GPUs have seen an order-of-magnitude increase in memory bandwidth with the adoption of newer generations of HBM. Meanwhile, GPU database systems are gaining traction, many building on cuDF, an open-source library of GPU relational operators. Previously, query performance was bound by memory bandwidth, but the increase in memory bandwidth has not resulted in a proportional speedup of cuDF kernels. To investigate why performance has not kept up, we built Valk, a performance analysis tool that combines data from multiple profilers. We profile cuDF running TPC-H in-memory on two extremes of hardware capability, the L4 and GH200 GPUs. The GH200 has 13.4$\times$ the memory bandwidth and 2.5$\times$ the instruction throughput of the L4, yet is only 5.2$\times$ faster in running TPC-H. Our analysis shows that when memory bandwidth is increased, kernels become compute bound. From our analysis, we make three recommendations to fully utilize the GPUs' potential for relational workloads when the memory wall is removed: kernels need to 1) make more efficient use of caches, and 2) increase occupancy and/or instruction level parallelism, and 3) execute fewer instructions per memory access. 
\end{abstract}

\renewcommand{\titlebreak}{\\}
\maketitle
\renewcommand{\titlebreak}{}

\ifdefempty{\vldbavailabilityurl}{}{
\vspace{.3cm}
\begingroup\small\noindent\raggedright\textbf{Artifact Availability:}\\
The source code, data, and/or other artifacts have been made available at \url{\vldbavailabilityurl}.
\endgroup
}

\section{Introduction}

GPUs are increasingly being used for relational analytics~\cite{RosenfeldSurvey2022,coddspeed,maximus,boss,siriusdb}. It makes sense to explore running analytics on GPUs given the abundance of GPU capacity, especially as AI workloads move to the latest generation of GPUs, leaving large pools of older GPUs underutilized.  GPUs offer orders of magnitude more bandwidth and compute than CPUs, but have been tuned and architected for AI workloads, which demand higher memory bandwidth~\cite{AIMemoryWallGholami2024, DataMovementIsAllYouNeedIvanov2021}. 

\begin{figure}[t!]
 \centering
\includegraphics[width=\linewidth]{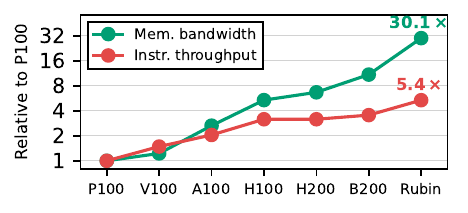}
 \caption{GPUs have increasingly more memory bandwidth compared to CUDA core instruction throughput.\protect\footnotemark}
 \label{fig:gputrend}
\end{figure}
\footnotetext{Clock frequencies for the B200 and Rubin are unpublished, we assume for those GPUs the clock frequency of the H200's.}

Intuitively, one would expect that the higher memory bandwidth of GPUs would also benefit relational workloads. Unfortunately, this is not the case. 
Although new datacenter GPUs have access to increasingly more memory bandwidth, the CUDA core instruction throughput of GPUs (those relevant for relational workloads) has not kept up. \autoref{fig:gputrend} plots the memory bandwidth and instruction throughput for different generations of GPUs showing the growing gap between the two. As a result of this gap, when running TPC-H, a GPU with 13.4$\times$ more bandwidth (GH200) delivers only 5.2$\times$ more performance (\autoref{tab:speedup}) than a smaller GPU (L4). For both GPUs the performance is roughly linear with the TPC-H scale factor, but the performance advantage of having more memory bandwidth is not correlated with the bandwidth ratios (and features as well as price) between the two GPUs. The reason for this is that the relevant instruction throughput is only 2.5$\times$ larger in the GH200 than in the L4. 
Previous work on earlier GPUs has suggested that query execution is memory bound (e.g.,~\cite{ShanbhagCrystal2020, CaoGPUDBCharacterization2023, Yuan2013}). In this paper, we study how the bottleneck has moved from the memory bandwidth to the \textit{CUDA core} instruction throughput, the instructions used by cuDF kernels. cuDF~\cite{cudf} is an open source library of relational GPU operators and the one most commonly used in GPU-based database systems \cite{siriusdb, maximus, prestodb, gqe, theseus, blazingsql}, thus representative of the state of the art in data analytics on GPUs. 

\begin{table}[t]
  \centering
  \setlength{\tabcolsep}{4pt}
  \caption{Instruction throughput, memory bandwidth, and the sum of kernel execution times, per GPU and SF. 
  }
\begin{tabular}{lrrr}
\toprule
\textbf{} & \textbf{L4} & \textbf{GH200} & \textbf{Ratio} \\
\midrule
Instr.\ throughput (G inst./s) & 413 & 1{,}037 & \textbf{2.5$\times$} \\
Mem. bandwidth (GB/s) & 300 & 4{,}022 & \textbf{13.4$\times$} \\
\midrule
TPC-H SF1 $\sum$ kernel (ms) & 84 & 34 & \textbf{2.4$\times$} \\
TPC-H SF3 $\sum$ kernel (ms) & 256 & 69 & \textbf{3.7$\times$} \\
TPC-H SF10 $\sum$ kernel (ms) & 924 & 204 & \textbf{4.5$\times$} \\
TPC-H SF30 $\sum$ kernel (ms) & 3{,}142 & 606 & \textbf{5.2$\times$} \\
TPC-H SF100 $\sum$ kernel (ms) & - & 2{,}020 & - \\
\bottomrule
\end{tabular}

	\label{tab:speedup}
\end{table}

To explore the behavior of cuDF in detail, we conduct an in-depth exploration and performance analysis of running relational workloads (TPC-H) using Maximus~\cite{maximus} and cuDF kernels. We focus on two points of the architectural space, a smaller, more economical GPU with limited memory bandwidth (NVIDIA L4), and a large, high end GPU with high memory bandwidth (NVIDIA GH200). 
To facilitate the exploration, we have built \textit{Valk}, an novel analysis tool that combines data from multiple profilers. 

The experimental exploration shows  that the 13.4$\times$ increase in memory bandwidth from the L4 to the GH200 is not matched with an equal increase in TPC-H performance. We use roofline analysis~\cite{roofline2009} to demonstrate that the kernels used in cuDF to run the key operators in TPC-H queries are mostly instruction bound instead of being memory bound. Furthermore, we show that the instruction throughput that is available is underutilized on both GPUs for compute bound kernels, indicating that much more work needs to be devoted to implementing relational operators on GPUs. We make three recommendations to alleviate this new bottleneck, recommendations that should help in developing a new generation of more efficient operators on GPUs that make better use of the GPU. These insights are also useful to inform the development of future GPUs oriented towards data analytics rather than AI workloads. We also identify which cuDF kernels make up more than 90\% of the workload on both GPUs, indicating which kernels are most affected by the new instruction bottleneck and those that should be redesigned first. 

The main contributions of the paper include:
\begin{itemize}
	\item Valk, a tool and visualization for performance analysis of queries on GPUs. We open source Valk including all experimental data.
	\item A roofline analysis that showing how cuDF kernels moved from being memory bound to compute bound.
	\item A performance analysis that identifies three sources of ineffiencies when the memory wall is lifted.
	\item A cuDF kernel analysis that identifes the most important kernels for executing TPC-H and how their performance shifts across the two GPUs.
\end{itemize}

\section{Background} \label{section:background}

We introduce relevant GPU concepts, the roofline model used to analyse kernel performance, and the cuDF library.

\subsection{GPUs} 

GPUs consist of \textit{streaming multiprocessors} (SM) divided into four \textit{subpartitions} (SMSP), each one with a \textit{scheduler} and multiple \textit{warps}. A warp consists of 32 \textit{threads}, and implements a Single Instruction, Multiple Threads execution model (SIMT~\cite{simt}). A warp issues one instruction for all threads in the warp. Every clock cycle, each scheduler picks one warp that may issue one instruction to the instruction pipelines on the SM. The maximum \textit{instructions per cycle} (IPC) per SM is therefore equal to the number of subpartitions. The achieved IPC may be lower, if in one clock cycle all warps in the subpartition are \textit{stalled}. Stalls occur if the warp needs to wait before issuing the next instruction, e.g., when waiting for an instruction dependency or memory latency. The IPC can be increased by either decreasing dependencies between instructions, i.e., increasing \textit{instruction level parallelism} (ILP)\cite{HennessyComputerArchitecture2017, ThesisVolkov2016, tuningdensealgebraVolkov2008}, or by increasing the number of active warps~\cite{cudabestpracticesguide}, i.e., increasing \textit{occupancy}.

The memory hierarchy of a GPU consists of one L1 cache per SM, and an L2 cache shared by all SMs, backed up by \textit{device memory}, the RAM of the GPU. The most fine grained access unit of device memory is a \textit{sector}, which consists of 32 consecutive bytes.

GPUs execute programs by launching \textit{kernels}. Kernels are programs executed over a large amount of threads, where groups of threads are subdivided into \textit{thread blocks}executed on a SM. A kernel might launch more thread blocks than there are SMs. In this case, the thread blocks are executed across multiple \textit{waves}. When there are multiple waves, later thread blocks need to wait until the previous thread blocks finish executing on the SM. A kernel might not launch enough thread blocks to launch a full wave, or the last wave is not full, and then the thread blocks do not span all SMs and the GPU is underutilized during that wave. 

\subsection{Roofline Model} 

The roofline model~\cite{roofline2009} is a tool for performance analysis that shows whether kernels are bottlenecked by memory bandwidth or instruction throughput. Canonically, the model consists of a graph with instructions/operations per byte of memory traffic on the x-axis, and instructions/operations per second on the y-axis. Two additional lines are drawn on the graph. The first line is a horizontal line that indicates the maximum theoretical instruction throughput. The second line is a diagonal line that crosses the first line and indicates the maximum memory bandwidth. These two lines form the \textit{roofline}. Additionally, a \textit{ridge line} can be drawn, which is a vertical line that crosses through the point where the diagonal memory bandwidth line and the horizontal floating point operations throughput line intersect. 

\begin{figure}[t!]
 \centering
 \includegraphics[width=0.9\linewidth]{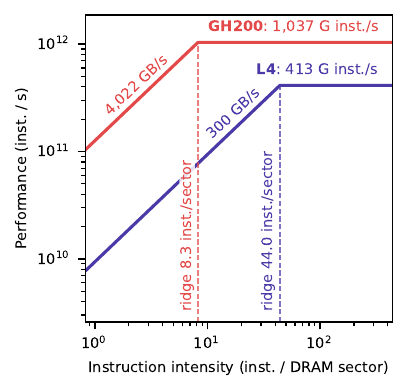}
 \caption{Roofline comparison of the L4 and GH200 GPUs.}
 \label{fig:roofline}
\end{figure}

Every kernel that is executed on the GPU can be plotted in this graph as a dot. If a kernel is located to the left of the ridge line, the kernel is \textit{memory bound}. Increasing the memory bandwidth will speed up this kernel. If a kernel is located to the right of the ridge line, the kernel is \textit{compute bound}. Increasing the instruction throughput will speed up the kernel. The kernel must be below either of the rooflines, as no kernel can achieve more than the theoretical limit. The closer a kernel is to this roofline, the more efficiently it utilizes the full potential of the hardware. 

In this paper, 
we use the instructions per second (instruction throughput) based roofline from Ding and Williams~\cite{instructionroofline2019}, created for workloads that only rarely execute floating point operations. In cuDF kernels, floating point operations are rare. In \autoref{section:roofline} we show that less than one percent of the issued instructions are floating point instructions. See \autoref{fig:roofline}  for an example for the L4 and GH200. 
This instruction throughput we consider in this paper is for warp instructions, the instructions that a warp issues for all threads active in the warp, and can execute an integer, memory, control, or similar operation. Instead of using \textit{operational intensity}, which is the number of floating point operations per DRAM byte, \textit{instruction intensity} is used, the number of (warp) instructions per DRAM sector. DRAM sectors are used instead of bytes, as sectors are the actual access unit in which memory is fetched from DRAM. For example, a random memory access during a join will fetch an entire sector from memory, even if only a few bytes are relevant.

For the memory bandwidth, we take the maximum memory bandwidth as reported by \textit{Nsight Systems}~\cite{nsightsystems}. To calculate the (warp) instruction throughput of a GPU, we use the number of streaming multiprocessors ($\text{N}_\text{SM}$), the number of subpartitions per streaming multiprocessor ($\text{N}_\text{SMSP per SM}$), and the measured clock frequency ($\text{CF}_\text{Measured}$). GPUs dynamically set their clock frequency and lower the clock frequency when needed to match thermal requirements. Therefore we do not use the theoretical max clock frequency, but measure the clock frequency during all 22 TPC-H kernels and all scale factors. We found the clock frequency to be stable, and therefore use the average. The instruction throughput is calculated as follows:

\begin{displaymath}
	\text{Inst. Throughput}=\text{N}_\text{SM} \times \text{N}_\text{SMSP per SM} \times \text{CF}_\text{Measured}
\end{displaymath}

This formula gives the number of warp instructions per second that a kernel can execute if all SMs on the GPU are active. It is important to note that the number of floating point operations per second that a GPU can execute is different from the number on warp instructions per second due to two reasons. 1) Warp instructions execute an instruction for all active threads within the warp, so a warp instruction can map to 32 operations, or more when the instruction performs multiple operations within an instruction, such as a fused multiply-add instruction. 2) GPUs use CUDA cores for most instructions, but also have access to \textit{tensor cores}, specialized floating point operations cores that can execute many more floating point operations per cycle per thread than even a fused multiply-add instruction~\cite{tensorcores}. Tensor cores are not used by cuDF. 

\subsection{cuDF} 

cuDF is an open source library of optimized relational database operators for GPUs originally developed for Data Frames. cuDF uses a columnar data layout compatible with Arrow~\cite{arrow}. Data processing workloads on GPUs with cuDF use the operator API. The operator API of cuDF calls one or more kernels, with each kernel completing a separate computation to execute the operator. These kernels can be small, e.g., when a single value needs to be computed and communicated to the CPU to decide the next step in the operator execution; or very large, when e.g., probing a build table as part of a join. The user of the API has no control over how and which kernels are launched. cuDF is used in many GPU database systems, such as SiriusDB~\cite{siriusdb}, Presto/Velox~\cite{prestodb}, GQE~\cite{gqe}, Theseus~\cite{theseus}, BlazingSQL~\cite{blazingsql}, and Maximus~\cite{maximus}. In \autoref{section:kernels}, we list the most important kernels making up more than 90\% of the aggregated kernel execution time and explain what they do. 

\section{Experimental Setup} \label{section:setup}

In what follows, we discuss the experimental setup and workloads used in the rest of the paper.

\begin{table}[t]
  \centering
  \setlength{\tabcolsep}{4pt}
  \caption{Comparison of L4 and GH200.}
\begin{tabular}{lrrr}
\toprule
\textbf{} & \textbf{L4} & \textbf{GH200} & \textbf{Ratio} \\
\midrule
Memory capacity (GiB) & 22 & 95 & 4.31$\times$ \\
L2 cache (MiB) & 48 & 60 & 1.25$\times$ \\
L1 cache / SM (KiB) & 128 & 256 & 2.00$\times$ \\
\midrule
SM count & 58 & 132 & 2.28$\times$ \\
SMSP per SM & 4 & 4 & 1.00$\times$ \\
Threads per SM & 1{,}536 & 2{,}048 & 1.33$\times$ \\
Base clock (GHz) & 0.80 & 1.53 & 1.92$\times$ \\
Boost clock (GHz) & 2.04 & 1.98 & 0.97$\times$ \\
Measured SM clock (GHz) & $1.78$ & $1.96$ & 1.10$\times$ \\
\midrule
\makecell[l]{Instr.\ throughput (G inst./s)} & 413 & 1{,}037 & 2.51$\times$ \\
Memory bandwidth (GB/s) & 300 & 4{,}022 & 13.42$\times$ \\
Roofline ridge (inst./sector) & 44.0 & 8.3 & 0.19$\times$ \\
\bottomrule
\end{tabular}

  \label{tab:hardwarecomparison}
\end{table}

\subsection{Hardware} 

\begin{table*}[t!]
  \centering
  \setlength{\tabcolsep}{4pt}
	\caption{Information each profiler provides, and the data Valk extracts from each. 
    }
	\definecolor{markgreen}{HTML}{1B7F3B}
\definecolor{markred}{HTML}{C0392B}
\definecolor{markamber}{HTML}{D68910}
\providecommand{\cmark}{\textcolor{markgreen}{\ding{51}}}
\providecommand{\xmark}{\textcolor{markred}{\ding{55}}}
\providecommand{\pmark}{\textcolor{markamber}{\ensuremath{\sim}}}
\providecommand{\ov}[1]{\textcolor{markred}{\textbf{#1}}}      
\providecommand{\nov}{\ensuremath{\cdot}}                       
\resizebox{\linewidth}{!}{%
\begin{tabular}{@{}l cc ccc cc l@{}}
\toprule
 & \multicolumn{2}{c}{\textit{Cost}} &  \multicolumn{5}{c}{\textit{Metrics}} & \textit{Role in Valk} \\
\cmidrule(lr){2-3} \cmidrule(lrlr){4-8} \cmidrule(lr){9-9}
\textbf{Profiler}
  & \makecell{Query\\overhead} & \makecell{Kernel\\overhead} 
	& \makecell{Query\\timing} & \makecell{Kernel\\timing} & \makecell{NVTX\\regions}
  & \makecell{System-wide\\metrics} & \makecell{Per-kernel\\counters} 
  & \makecell[l]{Data} \\
\midrule
None
  & \nov & -- 
  & \cmark & \xmark & \xmark
  & \xmark & \xmark 
  & baseline, validation \\
NSYS
  & \ov{+} & \nov 
  & \xmark & \cmark & \cmark
  & \cmark & \xmark
  & timeline, system metrics \\
NCU (NC)
  & \ov{++} & \nov 
  & \xmark & \cmark & \xmark
  & \xmark & \cmark 
  & time-sensitive counters \\
NCU (D)
  & \ov{+++} & \ov{++} 
  & \xmark & \xmark & \xmark
  & \xmark & \cmark
  & time-insensitive counters \\
\bottomrule
\end{tabular}
}

  \par\smallskip
	{\footnotesize \cmark~provided, \xmark~not available or distorted. Overhead: \nov~none, \ov{+}/\ov{++}/\ov{+++}~low/moderate/high; -- not measurable. \textit{NSYS}=Nsight Systems, \textit{NCU (NC)}=Nsight Compute with no clock control, \textit{NCU (D)}=Nsight Compute with default settings.}
  \label{tab:profilercomparison}
\end{table*}

We benchmark TPC-H on an L4 GPU and on a single GPU from a GH200. The GH200 system consists of a Grace CPU and one to four Hopper architecture GPUs analogous to the H100 SXM GPU~\cite{GH200Hoefler2024}. As the GPU in the GH200 does not have an individual name, we refer to the single GPU as ``GH200''.
We chose the L4 and GH200 because they differ substantially in price and memory bandwidth while remaining otherwise comparable as 
their underlying architecture (Ada Lovelace and Hopper, respectively) belongs to the same hardware generation~\cite{GH200Hoefler2024}. This makes them ideal for comparing kernel performance with low and high memory bandwidth.

\autoref{tab:hardwarecomparison} shows the main hardware properties of both GPUs, and the difference between them as a ratio. The GH200 has both more device memory capacity as well as larger caches for both L1 and L2. The GH200 has 13.4$\times$ more memory bandwidth but only 2.51$\times$ more cuDA core instruction throughput. \autoref{fig:roofline} shows how this disproportionate increase causes the roofline ridge line to shift from a kernel capable of executing 44 instructions per sector in the L4 before the kernel becomes compute bound, to being able to execute 8.3 instructions per sector in the GH200 before becoming compute bound. A kernel that is located at the ridge line on the GH200  can execute 5.3$\times$ fewer instructions per memory access than in the L4 to avoid becoming compute bound. 

The GH200 has 2.51$\times$ more instruction throughput than the L4, mostly due to the 2.28$\times$ increase in SMs, as the clock frequency is only slightly higher on the GH200, and the number of SMSPs of both GPUs is equal. The GH200 has more threads per SM, but the same number of SMSPs. On the L4, the scheduler has fewer warps to select instructions from, but because the number of SMSPs is equal, the number of instructions per cycle that can be executed per SM is equal on both GPUs. Because the higher instruction throughput is due to the increase in SMs, kernels find it harder to exploit the full throughput on the GH200 than the L4, as kernels need to launch more thread blocks to create at least one full wave of thread blocks.
The question we answer in the rest of the paper is how these hardware characteristics affect cuDF kernels, and thus the performance of database engines relying on cuDF. 

\subsection{Workloads and Execution}

We benchmark TPC-H using Maximus on scale factors 1, 3, 10, and 30. Scale factor 100 goes out of memory on the L4 for most queries, so it is not included in the comparison. However, we do publish scale factor 100 results for the GH200 in our artifacts and in Valk. We aim to measure how kernel performance differs across GPUs, without including any overhead from the database system that uses cuDF. Therefore, we define the query execution time as the sum of kernel execution times. Any overhead in creating query plans, or in executing the query plan from the CPU is ignored, as it can vary greatly per system and highly depends on the implementation. We also assume that all data is resident on the GPU before the query starts, as the interconnect overhead depends on the interconnect used, and how well the system can overlap query data transfer with query execution. The role of the interconnect has already been extensively studied in previous work~\cite{maxbench, Yuan2013, DataPlacementYogatama2022, CoprocessingHe2013, PumpUpTheVolumeLutz2020}. 
We execute each query five times, and take the measurements of the last repetition. We use cuDF version 25.12 and compile and run everything in a Docker image (cuda:13.0.3-cudnn-devel-ubuntu24.04) to ensure that the environments are identical across both GPUs.

\section{Valk} \label{section:valk}

As part of our analysis efforts, we have built \textit{Valk}. Valk is an open source tool for studying analytics on GPUs. It works as a meta profiler, and combines data from multiple profilers and workload information from Maximus into a unified dataset. The dataset is structured as a set of Parquet~\cite{parquet} files. 
All results reported in this paper are based on this unified dataset. Valk offers a user interface that displays the combined data from the profilers and engine specific information. It shows  timelines, roofline graphs, queryable tables, bar charts, and query plans, giving the user a comprehensive overview on how the query is executed on the GPU.
Valk is a useful tool for understanding query execution on GPUs but also provides crucial insights on how to do performance analysis on a GPU. 

\subsection{Dataset}

\parheader{Data sources.} Currently the tool supports the following data sources: execution traces from Maximus, system profiling data from NVIDIA Nsight Systems~\cite{nsightsystems}, and per kernel hardware counters data from NVIDIA Nsight Compute~\cite{nsightcompute}. Each of these data sources contain different slices of information on how a query was executed on the GPU. Individually, none of these tools provides a complete view of the execution of a query, motivating the need for a tool that combines the information from all profilers.

\autoref{tab:profilercomparison} summarizes the execution overhead, measured metrics, and information ingested by Valk for each profiler. Besides running queries with a profiler active, Valk also runs the query without any profiler active. This run is used to measure the query execution time without any profiling overhead. Then Valk runs the query once with Nsight Systems, and two times with Nsight Compute with different settings active. 

\parheader{Nsight Systems.} Nsight Systems is a low-overhead profiler that returns detailed execution information such as timestamps of all kernels, memory copies, and NVTX regions~\cite{nvtx}. These timestamps can be used to construct a timeline of all GPU events. NVTX regions are a method to annotate code regions with semantic information, and is used in Valk to detect which operator is active at any point in time. 

\parheader{Nsight Compute.} Running a query with Nsight Compute imposes a high overhead on the end-to-end query execution time, but returns detailed kernel execution information using hardware counters. The query is executed twice with Nsight Compute, once with default settings, and once with clock control disabled. Nsight Compute will repeat each kernel many times in order to collect a large set of metrics. Each repetition collects a different subset of metrics. In order to combine the data from multiple kernel repetitions, Nsight Compute fixes by default the clock frequency to the base frequency, as this avoids distortions from different repetitions running with different clock frequencies. However, executing kernels with a clock fixed to base clock frequency lowers the instruction throughput, increasing the kernel execution time, and distorts time sensitive measurements such as executed instructions per second. Additionally, the GPU's memory clock is separate from the GPU's main clock, and runs at a separate frequency that is constant (and is not changed by Nsight Compute). This distorts the roofline of kernels run with a fixed base clock, as the instruction throughput is lower, but the memory bandwidth is the same. Therefore we run the query with Nsight Compute a second time with clock control disabled, and collect only a limited set of time-sensitive metrics that can be collected within a single repetition. This bypasses the constraint that, for multiple repetitions, the clock frequency must be constant.

\parheader{Combining Profiler Data.} Valk combines the data by validating that each profiler executed the query identically and then matches all data from different profilers to stitch together a complete view of the query's performance. We instrumented Maximus with more detailed logging to log query plans and operator statistics, such as row counts, selectivity, and match ratios. These logs form an execution trace that can be compared across query executions by different profilers to ensure each profiler executed the query in the same way. Then we match data from Nsight Systems with Nsight Compute by comparing the profiled kernels from both, and if the count, order, and names match, we can combine the data from both profilers. In our dataset Valk was able to match the profiler data for almost all queries, except for Q2 at SF3 \& Q11 at SF 1 and 3 on GH200, and Q11 at SF 1 on the L4. These missing queries appeared at low scale factors and only account for 10 milliseconds of execution time out of a total of 7.9 seconds of execution for all sacle factors, queries, and GPUs, resulting in less than 0.2\% of total time not being matched. They are excluded from the dataset.

\parheader{Profiler Overhead.} In \autoref{tab:profiling-overhead} the overhead of each profiler is shown, split in end-to-end query execution overhead and kernel timing overhead. This data shows that using Nsight Systems to construct a timeline is possible, as the end-to-end query overhead is minimal. The overhead is slightly larger for smaller scale factors than larger scale factors, indicating that NSYS has a fixed overhead cost per measurement that is amortized more effectively on larger scale factors. On the GH200 the overhead is larger, as all queries are processed faster, so the fixed overhead cost distorts the timings more. The overhead of Nsight Systems on individual kernels can not be measured, as running without any profiler can not measure individual kernel times. Nsight Compute with default settings can not be used for time-sensitive metrics, as the fixed base clock distorts the kernel execution times. Nsight Compute with clock control disabled matches the kernel times of Nsight Systems closely. Nsight Systems' measurements are considered the source of truth for kernel execution times as the end to end query overhead is smaller, and the timeline depends on Nsight Systems' timestamps.

\subsection{User Interface}

\begin{table}[t!]
\centering
    \caption{Ratio of end-to-end query execution times in comparison to no profiler, and ratio of sum of kernel execution times in comparison to Nsight Systems' measurement. Individual kernel execution times are unknown without a profiler.}
\resizebox{\linewidth}{!}{%
\begin{tabular}{lrrrr}
\toprule
 & \multicolumn{2}{c}{L4} & \multicolumn{2}{c}{GH200} \\
\cmidrule(lr){2-3} \cmidrule(lr){4-5}
\textbf{Profiler} & \textbf{Query} & \textbf{Kernel} & \textbf{Query} & \textbf{Kernel} \\
\midrule
None & 1.00$\times$ & -- & 1.00$\times$ & -- \\
NSYS & 1.04$\times$ & 1.00$\times$ & 1.14$\times$ & 1.00$\times$ \\
NCU (NC) & 7.57$\times$ & 0.96$\times$ & 126$\times$ & 1.01$\times$ \\
NCU (D) & 3,057$\times$ & 1.36$\times$ & 4,989$\times$ & 1.21$\times$ \\
\bottomrule
\end{tabular}
}

\par\smallskip
    {\footnotesize\textit{NSYS}=Nsight Systems, \textit{NCU (NC)}=Nsight Compute with no clock control, \textit{NCU (D)}=Nsight Compute with default settings.}
\label{tab:profiling-overhead}
\end{table}

The user interface of Valk is based on three pages. The first page is the \textit{matrix page}, which shows an overview of the execution times and relative speedup of GH200 over L4 of all queries across all scale factors (including SF100 for GH200) to identify interesting queries to analyse. After query selection, the user is directed to the second page. The second page is the core performance analysis page, the \textit{inspector page}. It contains interactive data visualizations, such as query timeline, roofline graphs, bar graphs, query plan, and queryable tables. The third page, the \textit{schema page}, gives users an explanation of the metrics shown in the inspector.

\begin{figure*}[t!]
      \centering
    \includegraphics[width=\linewidth]{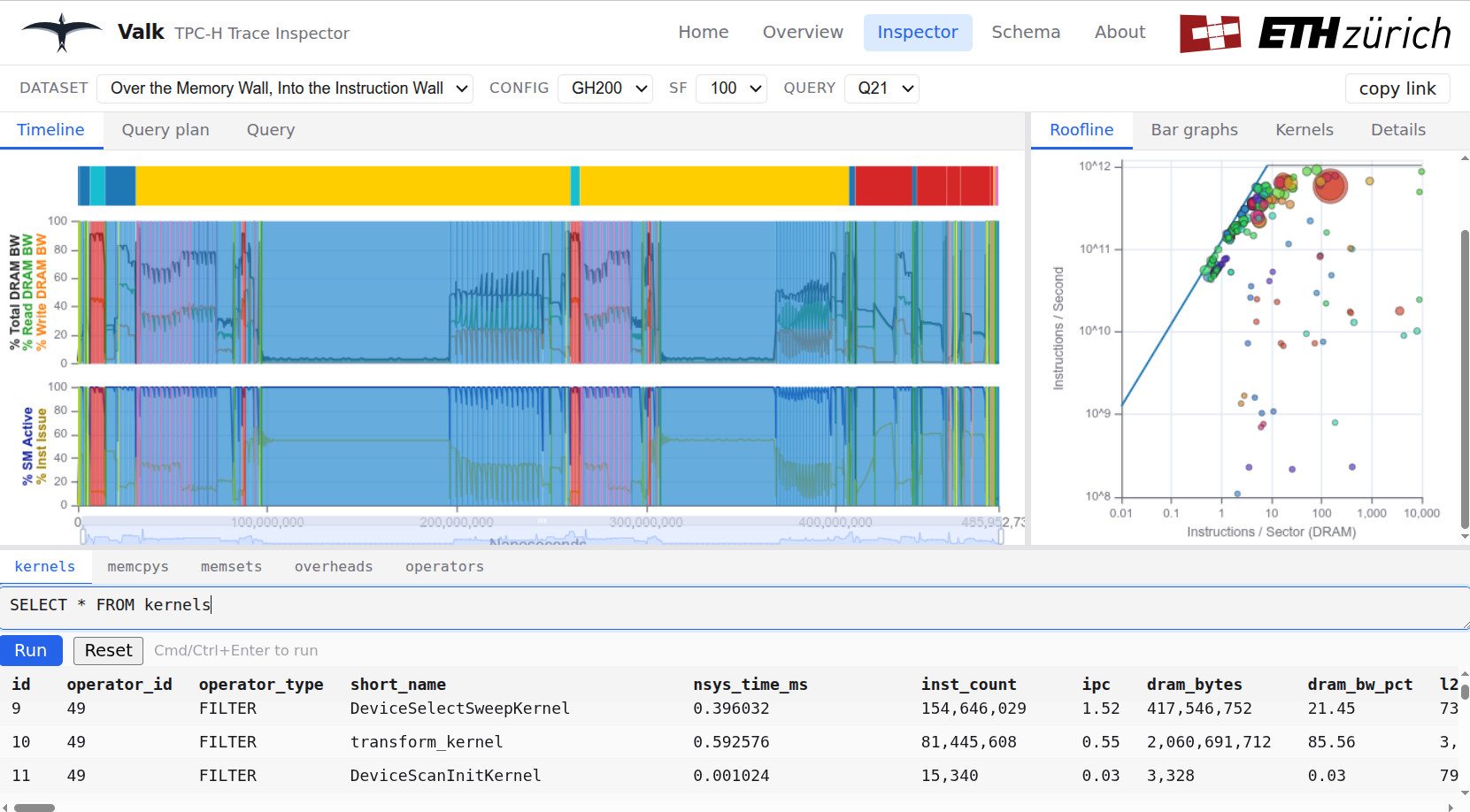}
    \caption{The inspector page of the Valk UI.}
    \label{fig:valk}
\end{figure*}

\parheader{Inspector overview.} The inspector page offers many different data visualizations as shown in \autoref{fig:valk}. The main visualization is the timeline view, which shows the duration and order of operators, kernels, and memory copies.  To the right of the timeline is the sidebar, which has four subtabs: roofline, bar graphs, kernels, and details. Below the timeline and sidebar is the metrics table. This table contains all metrics that the visualizations are based on, and enables the user to inspect all data interactively.

\parheader{Timeline.} The timeline has three bars: the upper bar is the operator bar, the middle bar is the memory bar, and the lower bar is the compute bar. The operator bar shows when each operator is active, and each operator type has a unique color. The memory and compute bar both show when kernels and memory copies are active. The memory bar is overlaid with a line graph of memory bandwidth usage, showing three lines for: read bandwidth percentage, write bandwidth percentage, and total bandwidth percentage. The compute bar is also overlaid with a line graph, one line for percentage of streaming multiprocessors active, and one line for percentage of instruction throughput. The timeline section also has a sub tab for showing the query plan.

\parheader{Sidebar.} The sidebar's main tab is the roofline graph, which shows each kernel as a circle, where larger circles indicate that that kernel was active for longer, and kernels that executed the same function share the same color. The sidebar also has tabs for three additional visualizations. 1) The bar graphs tab shows for each kernel the distribution in instructions per pipeline, and the distribution of stall reasons. 2) The kernels tab shows for a selection of multiple kernels, how much time is spent percentage wise in each distinct kernel, and how many times each distinct kernel is executed. 3) The details tab shows summary metrics depending on the current selection.

\parheader{Table.} The metrics table in the lower half of the inspector shows the full dataset that all other visualizations are based upon. For every kernel, a wide range of metrics is shown. The table is interactive, clicking on a row in the table will highlight that kernel in the timeline, clicking on a column header will sort the table by that column, and \texttt{ALT} clicking a column header will show summary statistics such as sum, mean and standard deviations, for that metric in the details tab. The table is fully SQL queryable, above the table is a text input in which users can enter SQL queries, and when those queries are executed the table will show the result of that query. The kernels in the query result are highlighted in the timeline as well. This queryability of the tables allows users to perform powerful data analysis, and relate hardware counter data from Nsight Compute to timeline data from Nsight Systems. 

\section{Roofline Analysis} \label{section:roofline}

We perform a roofline analysis on all cuDF kernels used at all scale factors of TPC-H to understand why cuDF's speedup on the GH200 GPU over the L4 GPU is limited. 
We show the kernel execution time distribution across memory and compute bounds for both GPUs. To illustrate how the distribution shifts, we show a heatmap of kernel execution time on the rooflines of each GPU.

\begin{table}[t!]
\centering
\caption{Instruction mix per GPU. Integer arithmetic forms the majority of instructions, floating point is rare.}
\begin{tabular}{lrr}
\toprule
\textbf{Instruction Class} & \textbf{L4} & \textbf{GH200} \\
\midrule
Integer Arithmetic & 61.80\% & 56.74\% \\
Control & 13.82\% & 12.38\% \\
Memory & 10.65\% & 13.21\% \\
Bitwise Operations & 4.72\% & 4.59\% \\
Uniform & 3.79\% & 7.61\% \\
Inter-thread Comm. & 2.09\% & 1.86\% \\
Conversion & 0.44\% & 0.42\% \\
Float Arithmetic & 0.29\% & 0.27\% \\
\midrule
Other & 2.40\% & 2.93\% \\
\bottomrule
\end{tabular}

\label{tab:instructionmix}
\end{table}

\parheader{Instruction Types.} \autoref{tab:instructionmix} shows the instructions issued by cuDF kernels when executing TPC-H. Less than one percent of the issued instructions are floating point instructions. A majority of the instructions are integer instructions, with control and memory instructions accounting for around a quarter of all instructions. 

\begin{table}[t!]
\centering
\caption{Percentage of per-GPU kernel time spent in memory- and compute-bound kernels when executing TPC-H, split by whether the utilization of the bottleneck reaches 75\%.}
\begin{tabular}{lrr}
\toprule
\textbf{Kernel Time} & \textbf{L4} & \textbf{GH200} \\
\midrule
\emph{Memory Bound} & \emph{74.1\%} & \emph{22.5\%} \\
\quad Memory Bandwidth $\geq$75\% & 42.5\% & 9.2\% \\
\quad Memory Bandwidth $<$75\% & 31.7\% & 13.4\% \\
\midrule
\emph{Compute Bound} & \emph{25.9\%} & \emph{77.5\%} \\
\quad Instruction Throughput $\geq$75\% & 3.5\% & 7.4\% \\
\quad Instruction Throughput $<$75\% & 22.4\% & 70.1\% \\
\bottomrule
\end{tabular}

\label{tab:kernelboundness}
\end{table}

\parheader{Memory vs Compute Bound.} \autoref{tab:kernelboundness} shows the distribution of kernel time per GPU across roofline classification: memory versus compute bound. Furthermore, we show how often memory bound kernels utilize less than 75\% of the available memory bandwidth, and how often compute bound kernels utilize less than 75\% of the available instruction throughput. We expressly choose 75\% as a rather low boundary, since a kernel that cannot reach even this value is certainly not bandwidth or throughput bound. On the L4, most time (74\%) is spent in memory bound kernels, of which a majority uses more than 75\% of memory bandwidth. In contrast, the GH200 spends the most time (78\%) in compute bound kernels, of which a majority utilize less than 75\% of the available instruction throughput.

\insight{New Bottleneck}{On a GPU with low memory bandwidth, kernels are memory bound. On a GPU with high memory bandwidth, the kernels become compute bound and fail to fully exploit the available instruction throughput.}

\begin{figure}[t!]
 \centering
 \includegraphics[width=\linewidth]{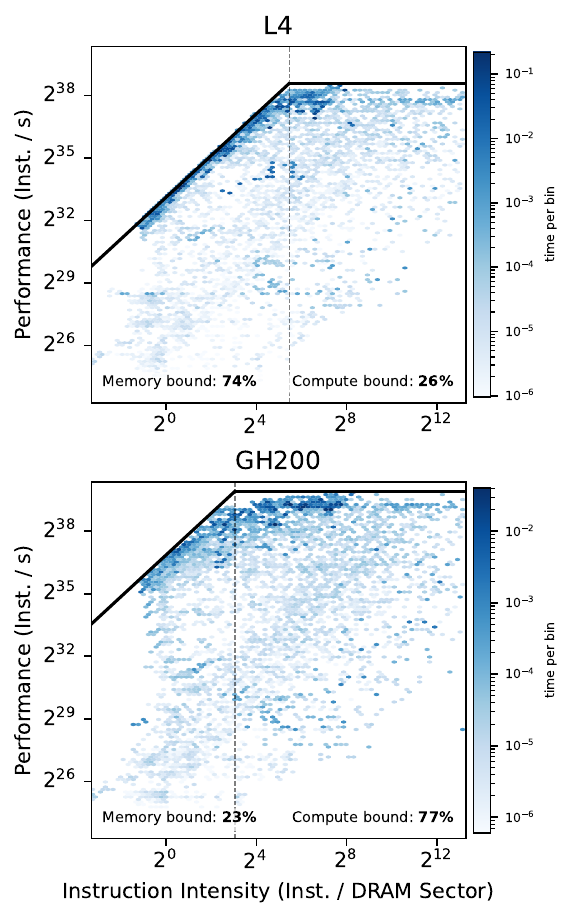}
 \caption{Logarithmic heatmap of kernel execution time plotted over a roofline graph of all cuDF kernels in TPC-H per GPU. Due to the larger increase in memory bandwidth than in instructions per second, the ridge line shifts, and more kernels become compute bound.}
 \label{fig:rooflineheatmap}
\end{figure}

\parheader{Causes.} To analyse the shift in bottleneck, we plot the execution time of all kernels on a combination of a roofline and a logarithmic heatmap (\autoref{fig:rooflineheatmap}). The heatmap shows the time distribution of all kernels in TPC-H up to and including SF 30 on a roofline plot. 
When we compare the heatmap of the L4 with the heatmap of the GH200, we observe two effects. 1) Due to the shifted ridge line, more kernels become compute bound.  2) Because of the larger increase in memory bandwidth, kernels that were memory bound experience relatively more speedup in comparison to compute bound kernels, so on the GH200 less time is spent on memory bound kernels.

\section{Performance Analysis} \label{section:analysis}

Due to the larger increase in memory bandwidth compared to the increase in instruction throughput, cuDF has become compute bound and the kernels are not fully utilizing the instruction throughput. Worse, now that cuDF is compute bound, cuDF will not benefit as much from the trend of GPUs being equipped with more memory bandwidth. Therefore it is imperative that cuDF kernels (and data processing kernels in general) account for this shift in hardware balance, make use of the available instruction throughput as much as possible, and optimize for the new compute bottleneck. 

In this section, we identify three recommendations on how to optimise cuDF kernels for the new compute bottleneck. Beforehand, we note that, at lower scale factors, kernels severely underutilize the GPU which lowers the individual kernel's performance, but is not problematic for the system as a whole the execution of those kernels can be overlapped although it is not clear that it makes sense to use a GPU for such small data sizes. 

We start the analysis by noting that the instructions per cycle (IPC) is low across all kernels, and we breakdown why by examining the stall reasons. We split the stall reason analysis into compute bound kernels and memory bound kernels. For compute bound kernels, we note that occupancy can be increased on both GPUs, but that even with ideal occupancy, peak IPC can not be reached in a large part of the kernels. Instruction-level parallelism (ILP) should also be increased. For memory bound kernels, we show that both the L1 and L2 cache do not achieve peak throughput in the large majority of kernels. Cache-aware algorithms could help reduce the average memory access latency, which would cause the number of stalls due to memory stalls to decrease as well. 

Finally, we show that even if stalls are perfectly overlapped and all kernels reach peak IPC, the majority of time on the GH200 would be spent in kernels that are instruction throughput bound. Those kernels will still not be exploiting the hardware trend of relatively more available memory bandwidth and relatively smaller increases in instruction throughput. Thus, even if all kernels achieve peak IPC, the speedup of GPU data processing would still largely be bound by increases in instruction throughput. Therefore we conclude that kernels need to decrease their instruction intensity to exploit the larger increases in memory bandwidth.

\subsection{Saturation}

\begin{figure}[t!]
  \centering
  \includegraphics[width=\linewidth]{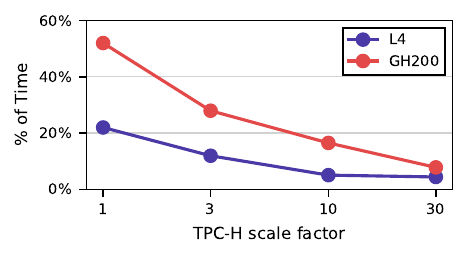}
	\caption{The percentage of time spent in a kernel that does not launch at least one full wave of threads.}
  \label{fig:kernelsaturation}
\end{figure}

At smaller scale factors, kernels tend to launch fewer thread blocks, and this results in more kernels that do not manage to launch even a single full wave of thread blocks. \autoref{fig:kernelsaturation} shows that at SF 1, 22\% and 52\% of kernel execution time on the L4 and GH200 respectively is spent in kernels that do not launch at least one full wave. As instruction throughput is the product of the number of SMs and clock frequency, not using all SMs diminishes the maximum instruction throughput kernels can reach. Undersaturation has  a larger effect on kernels on the GH200, because the increase in instruction throughput over the L4 is largely due to the higher SM count, as the clock frequency is approximately equal. The effect is reduced at large scale factors, but underutilization might occur when using several GPUs in parallel to process a query over skewed data. 

\insight{Undersaturation}{Kernels find it harder to exploit the full instruction throughput of GPUs with many SMs, especially when data sizes are small. }

We do not classify these undersaturated kernels as problematic, as the database using cuDF can use multiple streams of kernels to run kernels concurrently. An individual kernel might not exploit the full instruction throughput, but the set of concurrent kernels as a whole could. Thus, kernels should be optimized to saturate GPUs with many SMs by exposing more data parallelism across SMs. 

\subsection{Instructions Per Cycle}
\begin{table}[t!]
\centering
\caption{Geometric mean IPC for memory bound and compute bound kernels on both GPUs. Compute bound kernels only achieve marginally more than half of maximum IPC.}
\begin{tabular}{lrr}
\toprule
\textbf{Kernels} & \textbf{L4} & \textbf{GH200} \\
\midrule
All Kernels & 0.96 & 1.95 \\
\midrule
Memory Bound & 0.58 & 0.90 \\
Compute Bound & 2.09 & 2.32 \\
\bottomrule
\end{tabular}

\label{tab:ipcboundness}
\end{table}

\autoref{tab:ipcboundness} shows that the IPC on both L4 and GH200 is 0.96 and 1.95 respectively, less than half of maximum IPC ($4.0 = \text{number of SMSP per SM}$). The IPC of memory bound kernels is lower than that of compute bound kernels, as memory bound kernel instructions are bound by memory bandwidth. However, even when only considering the IPC of compute bound kernels, the IPC is less than 60\% of maximum IPC on both GPUs.

\insight{Low IPC}{The instructions per cycle for compute bound kernels is low on both GPUs.}

\begin{figure}[t!]
  \centering
  \includegraphics[width=\linewidth]{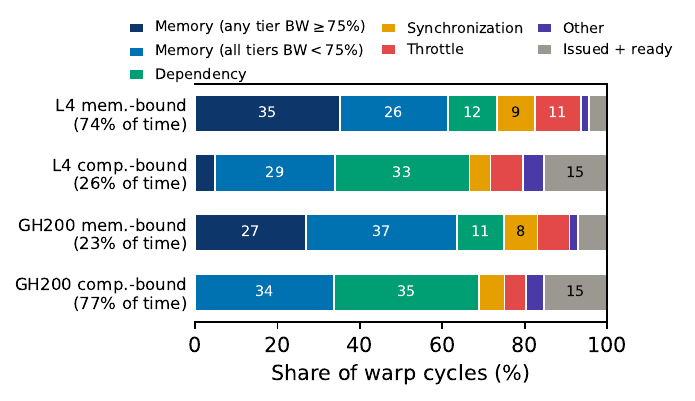}
	\caption{Stall breakdown per GPU and kernel bound. Memory stalls are split up into stalls due to throughput (a kernel achieving more than 75\% bandwidth on L1, L2, or main memory) and stalls due to latency (a kernel achieving less than 75\% bandwidth on L1, L2, and main memory).}
  \label{fig:stallbreakdown}
\end{figure}

\parheader{Stalls.} To investigate the reasons for low IPC, we show a breakdown of stall reasons per GPU and split in memory and compute bound kernels in \autoref{fig:stallbreakdown}. As expected, the number of memory stalls is higher in memory bound kernels, on both GPUs. We split memory stalls up per kernel, based on whether that kernel achieved more than 75\% bandwidth on any tier of the memory hierarchy (L1, L2, main memory). This distinction is made because kernels that saturate bandwidth at any tier will have more stalls stalled on memory requests delayed due to throughput constraints. However, a kernel that does not saturate the bandwidth of any tier in the memory hierarchy and encounters a memory stall is stalled due to the latency of the memory request. In memory bound kernels, many kernels are still bound by memory latency related stalls. This is a result of memory bound kernels not always maximizing memory bandwidth, as was shown in \autoref{tab:kernelboundness}.

\parheader{Compute Bound Stalls.} Compute bound kernels encounter low IPC due to 1) memory latencies while not saturating throughput on any memory tier, and 2) instruction dependencies. These two categories cause 60-70\% of all stalls for compute bound kernels. Both categories can be mitigated by overlapping latencies by either increasing occupancy or by increasing instruction level parallelism (ILP)~\cite{HennessyComputerArchitecture2017, tuningdensealgebraVolkov2008, ThesisVolkov2016}. Increasing occupancy hides latencies by having more warps active that can issue instructions, at the cost of fewer memory resources per warp~\cite{cudabestpracticesguide}. With instruction level parallelism, instructions have fewer interdependencies, allowing warps to be able to issue a new instruction sooner. GPU kernels might draw inspiration from techniques such as the vectorized query execution model, which unrolls loops and uses branchless code that exposes large amounts of ILP to the compiler, which in turn increases IPC on CPUs~\cite{HyperPipeliningQueryExecutionboncz2005, Zukowski2006VectorizedHashing, Zukowski2006VectorizedCompression}.

\begin{figure}[t!]
  \centering
  \includegraphics[width=\linewidth]{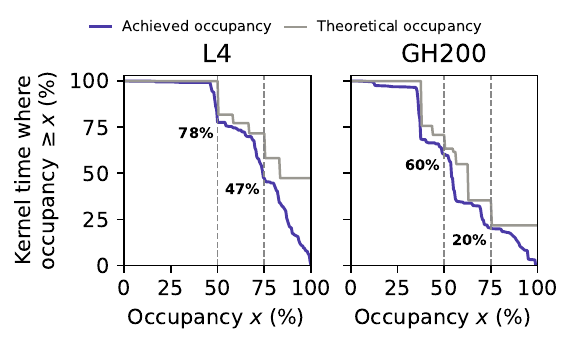}
	\caption{Occupancy is low on the GH200, and can still be increased on the L4.}
  \label{fig:occupancy}
\end{figure}

\parheader{Occupancy.} The percentage of warps that are active is called the \textit{occupancy}, which can be increased by reducing the number of resources, such as registers, that each warp uses. If there are more warps active, latencies can be hidden by overlapping work. The total number of resources on a SM is fixed, and the fewer resources an individual active warp uses, the more warps can be active simultaneously. \textit{Theoretical occupancy} indicates how many warps can be active at the same time, while \textit{achieved occupancy} shows how many warps were active averaged over all cycles a kernel was active. The achieved occupancy is smaller than theoretical occupancy, as some warps finish their work earlier than other warps.  

\autoref{fig:occupancy} shows how much kernel time is spent at each level of occupancy. The occupancy is higher on the L4, as the L4 has fewer warps per SM, but more registers per warp. This decreases the register pressure on L4 SMs, and enables more occupancy. However, on both GPUs a majority of kernel time is spent in kernels with 75\% occupancy or less, showing that kernels could be optimized to increase occupancy and hide more latency by having more warps active concurrently.

\insight{Low Occupancy}{Occupancy on both GPUs has room for improvement, showing that kernels could hide more latency with overlapping more work by increasing occupancy.}

\begin{table}[t!]
\centering
	\caption{Share of kernel time reaching at least 2, 3 or 4 instructions per cycle (IPC), counting only kernels that stay below 75\% of L1, L2 and memory bandwidth. In the scaled column every kernel's occupancy is scaled to 100\%, where IPC grows linearly in the added warps. Even under that ideal occupancy, only 51.1\% of GH200 kernel time reaches peak IPC, and none of L4's.}

\resizebox{\linewidth}{!}{%
\begin{tabular}{lrrrr}
\toprule
 & \multicolumn{2}{c}{L4} & \multicolumn{2}{c}{GH200} \\
\cmidrule(lr){2-3} \cmidrule(lr){4-5}
\textbf{IPC (\% of peak)} & \textbf{Measured} & \textbf{Scaled} & \textbf{Measured} & \textbf{Scaled} \\
\midrule
2.0 (50\%) & 23.3\% & 55.4\% & 67.1\% & 88.1\% \\
3.0 (75\%) & 6.1\% & 25.7\% & 8.6\% & 76.2\% \\
4.0 (100\%) & 0.0\% & 0.0\% & 0.0\% & 51.1\% \\
\bottomrule
\end{tabular}
}

\label{tab:ilp}
\end{table}

\begin{figure*}[t!]
  \centering
  \includegraphics[width=\linewidth]{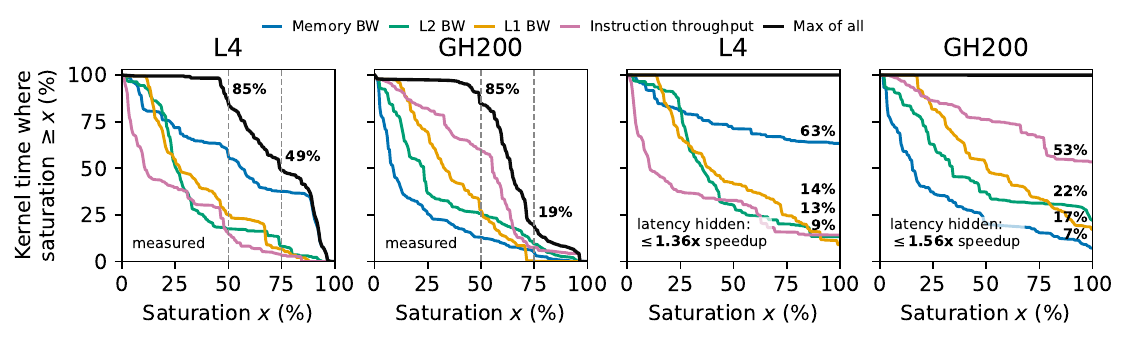}
	\caption{The left plots show the measured saturation across kernel time for instruction throughput and the utilized bandwidth of L1, L2 and memory. The right plots show saturation when IPC is increased until at least one of the resources reaches 100\% saturation. The max of all line shows the highest saturation on any resource by kernel time. }
  \label{fig:saturation}
\end{figure*}

\begin{figure}[t!]
  \centering
  \includegraphics[width=\linewidth]{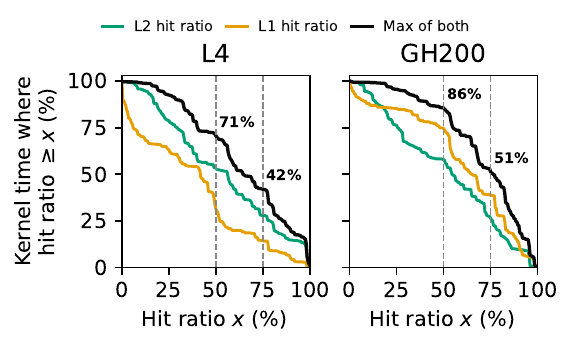}
	\caption{Cache hit ratios by kernel time and GPU.}
  \label{fig:cachehitratio}
\end{figure}

\parheader{Instruction-level parallelism.} \autoref{tab:ilp} shows the measured IPC and the IPC when all kernels would reach 100\% occupancy (disregarding any resource limits), assuming that IPC is increased linearly with the added active warps, and IPC is not bound by any memory tier. Kernels that utilise more than 75\% bandwidth of L1, L2 or memory are excluded because IPC is not relevant for those kernels, and because their IPC is lower as they are bound by bandwidth. This would skew the IPC to be lower while not being relevant for those kernels. The table shows that even if kernels would achieve ideal occupancy, they would still not saturate instruction throughput and reach a high IPC. Therefore \autoref{tab:ilp} shows that kernels need to increase their ILP, as only increasing occupancy would not achieve peak IPC. Also, kernels could consider increasing their ILP at the cost of occupancy, as this trade off is a performance optimization technique that can result in more performant kernels if the benefit of increasing ILP is higher than the cost of decreasing occupancy. Decreasing occupancy gives individual warps more resources, such as registers, which allows kernels to keep more data in registers, which can decrease memory spilling, enable different algorithm implementations, and increase opportunities for ILP~\cite{tuningdensealgebraVolkov2008, ThesisVolkov2016}.

\insight{Low Instruction-Level Parallelism}{Increasing occupancy alone would not result in peak IPC, showing that kernels need to consider other methods such as increasing instruction level parallelism as well.}

\recommendation{Overlap Latency}{To increase IPC, kernels need to hide latencies by overlapping more work by increasing occupancy and/or instruction-level parallelism.}

\subsection{Resource utilization} \autoref{fig:saturation} shows the resource utilization across L1, L2, \& memory bandwidth and instruction throughput by kernel time. The figure shows that the L4 achieves the highest resource utilization on memory bandwidth, while the GH200 utilizes instruction throughput the most, in line with what \autoref{fig:rooflineheatmap} showed. The 'Max of both' line shows the maximum resource utilization of across all of the four resources by kernel time. L4 kernels utilize at least 75\% of throughput of at least one resource 49\% of the time, while kernels on the GH200 only use more than 75\% of throughput of any resource 19\% of the time. This shows that on the GH200, 81\% of the time kernels are not throughput bound on any of the four resources.

\parheader{Caches} Stalls due to memory when none of the tiers are not throughput bound can be reduced by decreasing the latency of memory accesses. \autoref{fig:saturation} shows that on both GPUs, very little kernel time utilizes the full throughput of the L1 or L2 cache. This underutilization is not due to low cache hit ratios, \autoref{fig:cachehitratio} shows that the cache hit ratios are reasonable. By using algorithms or kernel implementations that more heavily rely on caches and exploit the underutilized cache throughput, memory latencies from main memory accesses can be reduced. 

\recommendation{Utilize Caches}{To decrease stalls due to memory latencies, kernels should increase cache utilization.}

\subsection{Instruction Intensity} 
Finally, \autoref{fig:saturation} shows that it is not enough to only increase IPC. \autoref{fig:saturation} shows in the right two plots what happens if kernels would be able to increase IPC up until at least one memory tier reaches 100\% utilization, or IPC is maximized. The point at which each resource line crosses the right border of the plot, indicates how much kernel time is bounded by that resource when IPC would be increased and the resourse utilization is linearly scaled by the same amount. Memory bandwidth would be the limiting factor for 63\% of kernel time on the L4, while only 9\% of kernels would be instruction throughput bound. 27\% of kernel time would be bounded by L1 or L2 cache throughput. For the GH200, 53\% of kernel time would be bounded by instruction throughput, while only 7\% of kernel time would be bounded by memory bandwidth. 39\% of kernel time would be bounded by L1 or L2 cache throughput. 

\begin{table*}[t!]
\centering
\caption{Key performance metrics of the 13 kernels that make up more than 90\% of the runtime on both GPUs at SF 30.}
\resizebox{\linewidth}{!}{%
\begin{tabular}{llrrrrllrrll}
\toprule
 &  &  &  & \multicolumn{4}{c}{L4} & \multicolumn{4}{c}{GH200} \\
\cmidrule(lr){5-8} \cmidrule(lr){9-12}
\textbf{Kernel} & \textbf{Op} & \textbf{\#Var} & \textbf{Speedup} & \textbf{\#} & \textbf{Sum} & \textbf{Mem.\%} & \textbf{IPC} & \textbf{\#} & \textbf{Sum} & \textbf{Mem.\%} & \textbf{IPC} \\
\midrule
\texttt{transform\_kernel} & JGFPO & 26 & 8.3$\times$ & 514 & 20.3\% & $81\!\pm\!27\%$ & $0.59\!\pm\!1.12$ & 515 & 12.1\% & $45\!\pm\!36\%$ & $1.90\!\pm\!1.57$ \\
\texttt{retrieve} & J & 2 & 2.8$\times$ & 49 & 10.7\% & $37\!\pm\!19\%$ & $1.79\!\pm\!0.47$ & 49 & 19.0\% & $7\!\pm\!7\%$ & $2.42\!\pm\!0.15$ \\
\texttt{DeviceMergeSortMergeKernel} & GO & 2 & 6.7$\times$ & 66 & 10.0\% & $54\!\pm\!40\%$ & $0.31\!\pm\!0.16$ & 66 & 7.4\% & $29\!\pm\!16\%$ & $0.94\!\pm\!0.55$ \\
\texttt{DeviceRadixSortOnesweepKernel} & GO & 2 & 12.3$\times$ & 32 & 8.2\% & $90\!\pm\!1\%$ & $0.20\!\pm\!0.09$ & 32 & 3.3\% & $71\!\pm\!9\%$ & $1.02\!\pm\!0.43$ \\
\texttt{DeviceMergeSortBlockSortKernel} & GO & 2 & 3.6$\times$ & 17 & 7.7\% & $13\!\pm\!4\%$ & $1.72\!\pm\!0.40$ & 17 & 10.5\% & $4\!\pm\!3\%$ & $1.89\!\pm\!0.65$ \\
\texttt{compute\_column\_kernel} & FP & 2 & 3.1$\times$ & 73 & 6.9\% & $71\!\pm\!21\%$ & $1.60\!\pm\!0.58$ & 73 & 10.9\% & $15\!\pm\!9\%$ & $2.28\!\pm\!0.29$ \\
\texttt{count} & J & 2 & 3.0$\times$ & 49 & 6.4\% & $59\!\pm\!17\%$ & $2.10\!\pm\!0.95$ & 49 & 10.5\% & $13\!\pm\!10\%$ & $2.64\!\pm\!0.25$ \\
\texttt{insert\_if\_n} & J & 2 & 10.5$\times$ & 58 & 6.1\% & $65\!\pm\!4\%$ & $0.33\!\pm\!0.16$ & 58 & 2.9\% & $42\!\pm\!11\%$ & $1.62\!\pm\!0.50$ \\
\texttt{static\_kernel} & JGO & 15 & 8.1$\times$ & 186 & 5.4\% & $85\!\pm\!12\%$ & $0.27\!\pm\!0.27$ & 192 & 3.3\% & $44\!\pm\!27\%$ & $0.94\!\pm\!0.40$ \\
\texttt{DeviceScanKernel} & JGFPO & 9 & 8.4$\times$ & 163 & 3.3\% & $85\!\pm\!13\%$ & $0.19\!\pm\!0.13$ & 163 & 2.0\% & $46\!\pm\!20\%$ & $0.67\!\pm\!0.33$ \\
\texttt{single\_pass\_shmem\_aggs\_kernel} & G & 1 & 3.4$\times$ & 10 & 3.3\% & $43\!\pm\!2\%$ & $1.22\!\pm\!0.02$ & 12 & 4.8\% & $9\!\pm\!1\%$ & $1.31\!\pm\!0.06$ \\
\texttt{gather\_chars\_fn\_string\_parallel} & JFO & 2 & 4.6$\times$ & 10 & 3.0\% & $75\!\pm\!4\%$ & $1.64\!\pm\!0.38$ & 10 & 3.3\% & $21\!\pm\!6\%$ & $2.82\!\pm\!0.07$ \\
\texttt{gather\_chars\_fn\_char\_parallel} & JGFO & 4 & 3.8$\times$ & 140 & 2.3\% & $80\!\pm\!22\%$ & $1.61\!\pm\!0.82$ & 140 & 3.0\% & $21\!\pm\!13\%$ & $2.74\!\pm\!0.62$ \\
\midrule
{\bfseries\boldmath Total} & {\bfseries\boldmath JGFPO} & {\bfseries\boldmath 71} & {\bfseries\boldmath 5.0$\times$} & {\bfseries\boldmath 1367} & {\bfseries\boldmath 93.7\%} & {\bfseries\boldmath $64\!\pm\!31\%$} & {\bfseries\boldmath $0.96\!\pm\!0.95$} & {\bfseries\boldmath 1376} & {\bfseries\boldmath 92.9\%} & {\bfseries\boldmath $22\!\pm\!24\%$} & {\bfseries\boldmath $1.99\!\pm\!0.90$} \\
\bottomrule
\end{tabular}
}

\par\smallskip
	{\footnotesize\textit{Op}=Operators that use the kernel: J=Join, G=Group By, F=Filter, P=Project, O=Order By, \textit{\#Var}=Number of templated variations, \textit{\#}=Number of invocations, \textit{Sum}=Sum of execution times as percentage of total, \textit{Mem.\%}=Percentage of maximum memory bandwidth used. \textit{IPC}=Weighted arithmetic mean and standard deviation of instructions per SM cycle.}
\label{tab:kernelbreakdown}
\end{table*}

\insight{Instruction Bottleneck}{Even if peak IPC is achieved, kernels on the high bandwidth GPU are bound by instruction throughput. This shows that with the current trend in GPU hardware, the bottleneck for data processing kernels is moving away from being memory bound to compute bound.}

\recommendation{Decrease Instruction Intensity}{Kernels should aim to execute fewer (warp) instructions per DRAM sector access to be able to exploit the large increases in memory bandwidth in newer GPUs.}

\section{Kernel Analysis} \label{section:kernels}

In this section, we analyse the GPU kernels that dominate aggregate execution
time across the 22 TPC-H queries on L4 and GH200. The 13 kernels listed
in~\autoref{tab:kernelbreakdown} account for more than 90\% of total kernel time on
both GPUs and are therefore load-bearing for TPC-H performance. Our goal is to
determine how their performance and bottlenecks shift between the
two GPUs. Overall, memory utilization on GH200 is only one third of
that on L4, as kernels become instruction-bound on the larger GPU.
We contextualize this trend by examining selected kernels in detail.

The most expensive kernel by aggregate time, and the one with the most variants,
is \texttt{transform\_kernel}, largely because of its high invocation count. It
applies a user-defined functor or lambda to the input element-wise and, through
indirection, also implements gather logic (\texttt{out[i]=in[map[i]]}) for
materialization. It therefore spans a wide range of instruction intensities and
access patterns, giving it the most diverse performance profile of any kernel we
study. Its backbone is identical across functors: a single unrolled main loop
that exposes more in-flight instructions. It does not, however, use vectorized
loads and stores, leaving memory bandwidth under-utilized. Its memory utilization
is lower on GH200 than on L4, implying that loop unrolling alone does not
saturate GH200's memory bandwidth. Use cases with frequent invocations and
sequential accesses should therefore consider specialized kernels with
vectorized loads and stores wherever possible.

The radix sort kernel (\texttt{DeviceRadixSortOneSweepKernel}) achieves the
highest memory utilization on both GPUs. It reads unsorted keys and values
sequentially and writes the sorted sequences back sequentially, using shared
memory as a write-combining buffer, and it unrolls aggressively to expose
instruction-level parallelism. As a result, GH200 outperforms L4 by a large
factor, yet it still reaches only 71\% memory utilization compared to 90\% on  the
L4. This mirrors the observation made for \texttt{transform\_kernel}:
sequential, high-ILP kernels that saturate L4's bandwidth leave a gap on GH200.

The \texttt{compute\_column\_kernel} evaluates arbitrary expression trees in cuDF
through an interpretation mechanism, which costs many additional instructions.
Despite its sequential memory accesses, it reaches only around 15\% memory
utilization on GH200, compared to 71\% on L4. This is a clear example of a
kernel turning from memory-bound to instruction-bound when moving from L4 to
GH200, and it suggests that on GPUs with abundant memory bandwidth, reducing
instruction count at the cost of additional sequential memory accesses can pay
off.

Two other kernels show low memory utilization on both GPUs, and lower
still on GH200. These kernels heavily rely on shared memory, which is located in the L1 cache, instead of main memory. \texttt{single\_pass\_shmem\_aggs\_kernel} aggregates values
per block in shared memory, avoiding a global memory update for every aggregated
value, and \texttt{DeviceMergeSortBlockSortKernel} likewise sorts in shared
memory. 

The hash join kernels \texttt{insert\_if\_n} (builds hash table), \texttt{count} (counts probe-side matches), and \texttt{retrieve} (materializes matches)
 read their input sequentially but probe the hash table
randomly. The \texttt{insert\_if\_n} kernel is memory bound on both GPUs, but suffers from long memory latencies into uncached rows in main memory. The \texttt{count} and \texttt{retrieve} kernels are memory bound on the L4, but become compute bound on the GH200 due to the ridge line shift limiting their speedup across the two GPUs.

For the rest of the listed kernels we only briefly explain what they do. The \texttt{static\_kernel} executes a \texttt{for\_each} loop over a range of indices using a user-supplied lambda. \texttt{DeviceScanKernel} computes a prefix sum over the results of a fused per-element functor, mainly to turn string sizes into a string column's offsets. The two \texttt{gather\_chars} kernels perform a \texttt{gather} operation on string columns to reorder them after an operator reordered the column.  

In summary, the most expensive TPC-H kernels use significantly less memory bandwidth
on the GH200 than on the L4. For kernels with sequential access patterns, high
instruction counts and non-vectorized accesses, both inconsequential on the L4,
become limiting on the GH200.

\section{Related Work} \label{section:relatedwork}
\parheader{CPU Performance Analysis.} Hardware-level performance analysis for query processing on CPUs is a well-studied subject. Boncz et al. studied how databases should be optimized for a main memory bandwidth bottleneck, as in-memory databases became feasible over disk-based databases~\cite{NewBottleneckBoncz1999}. Ailamaki et al.~\cite{WhereDoesTimeGoAilamaki1999} investigated where in-memory databases were leaving performance on the table by breaking down the reasons behind stalls of the CPUs, and proposing optimizations to reduce stalls. Later, Boncz et al. showed how instead of a Volcano-style execution model~\cite{VolcanoGraefe1994}, a vectorized execution model could reduce stalls and increase IPC by increasing instruction-level parallelism~\cite{HyperPipeliningQueryExecutionboncz2005}. Kersten et al.~\cite{EverythingCompiledVectorizedKersten2018} compared how vectorized execution and compiled query execution~\cite{CompiledQueriesNeumann2011} differ in performance and usage of the CPU.

\parheader{GPU Performance Analysis.} Previously, GPU performance analysis has mostly focused on SSB queries~\cite{ssb} and prototype operator implementations, and do not consider performance across GPUs with large differences in memory bandwidth and instruction throughput balance. Yuan et al.~\cite{Yuan2013} predict SSB query performance as implemented by their own restricted set of operators by creating a model that calculates the amount of data accessed and assumes that all queries are memory bandwidth bound. Shanbhag et al.~\cite{ShanbhagCrystal2020} create a model with the same assumptions, for their own tile-based operator implementations. Cao et al.~\cite{CaoGPUDBCharacterization2023} predict SSB query performance of multiple GPU database systems with a roofline based model. They predict performance when a GPU is split into multiple subunits using MIG~\cite{mig}, where each subunit has a similar balance of memory bandwidth and instruction throughput. Furthermore, there are numerous studies on the role of CPU-GPU and GPU-GPU interconnects for query processing~\cite{maxbench, Yuan2013, DataPlacementYogatama2022, CoprocessingHe2013, PumpUpTheVolumeLutz2020, Beischl2026AppleSilicon,distributed-tqp} that complement our study of the performance of cuDF kernels.

\parheader{Alternative Implementations.} There are also GPU database systems that do not use cuDF and implement operators themselves: TQP~\cite{tqp1, tqp2} and its successor Coddspeed~\cite{coddspeed} are closed source and build operators on top of PyTorch~\cite{pytorch}, BOSS~\cite{boss} builds operators on top of ArrayFire~\cite{arrayfire}, HeavyDB~\cite{heavydb} compiles queries into kernels with LLVM~\cite{llvm}, and Beischl et al.~\cite{Beischl2026AppleSilicon} implement a subset of SSB queries with their own operator implementations for Apple general purpose GPUs. Additionally, there is a significant amount of studies that do not implement a complete GPU database system, but implement, optimise and analyse a specific operator~\cite{join1, join2, join3, join4, join5, join6, join7, join8, joingroupby1, joingroupby2, filter1, groupby1, groupby2, groupby3, groupby4, sort1, sort2,wu2026eigerefficientlibrarygpubased}.

\parheader{Database Performance Analysis Tools.} To the best of the authors' knowledge, Valk is currently the only tool that combines detailed GPU kernel performance data with database specific context data such as query plans and operator statistics. Beischl et al.~\cite{Beischl2021Profiling, Beischl2025UmbraPerf} created a profiler for Umbra~\cite{umbra} that combines detailed performance data with database specific context, but Umbra is a CPU-based system. Quent~\cite{quent} is an experimental framework that can combine telemetry data from GPUs and database specific context, but does not offer detailed performance data on kernels. Outside of the database domain, there is a wide variety of performance analysis tools for GPUs that provide different data and insights, but none of them link the performance data with database specific context~\cite{nsightcompute, nsightsystems, hpctoolkit, gpa, drgpum}. Valk can expand it's data sources by directly using the underlying libaries of these related tools, such as CUPTI~\cite{cupti} or NVML ~\cite{nvml}.

\section{Conclusion} \label{section:conclusion}

In response to the increasing importance of AI workloads, GPUs have increased memory bandwidth in comparison to CUDA core instruction throughput. We show that TPC-H performance does not keep up with memory bandwidth, and use Valk, our performance analysis tool, to investigate why. We show that on a low-bandwidth GPU, cuDF kernels are memory bound, but on a high bandwidth GPU cuDF kernels become compute bound and do not fully exploit the available instruction throughput. We then demonstrate that throughput utilization could be improved by increasing occupancy and/or instruction-level parallelism to overlap work and hide latencies, and that cuDF kernels can utilize caches more efficiently to decrease memory latencies. Finally, we suggest that, even if latencies are completely hidden and peak instruction throughput is reached, cuDF kernels are still instruction throughput bound on the GH200 and will not benefit from the current trends in GPU hardware. Therefore, the number of warp instructions per sector access needs to be decreased as well. These are important insights to improve the implementation of relational operators on GPUs and to inform the architecture of GPUs to better support data analytics. 

\begin{acks}
	This work has been supported by a grant from the Swiss National Supercomputing Centre (CSCS) under project ID sm94 on Alps, and an unrestricted research donation from NVIDIA. 
\end{acks}

\bibliographystyle{ACM-Reference-Format}
\bibliography{sample}

\end{document}